# Blockchain in Space Industry
# - Challenges and Solutions-

**Mohamed Torky, Tarek Gaber, and Aboul Ella Hassanien**

**Scientific Research Group in Egypt (SRGE), Cairo, Egypt**

www.egyptsceince.net

**Abstract:** Blockchain technology can play a vital role in the space industry and exploration. This magic technology can provide decentralized and secure techniques for processing and manipulating space resources as *space digital tokens*. Tokenizing space resources such as orbits, satellites, spacecraft, orbital debris, asteroids, and other space objects in the form of blockchain-based digital tokens will reflect plenty of various applications in the space mining industry. Moreover, Blockchain algorithms based on smart contracts can be utilized for tracking all space transactions and communications in a transparent, verifiable, and secure manner. This paper is one of the first attempts towards conceptually investigating adopting blockchain theory in the space industry based on *space digital token* concept. A new conceptual blockchain in space industry framework is proposed, and new models are created for introducing proposed solutions for some major challenges in the space industry and exploration. Finally, the paper is ended with discussing *SpaceChain*, the first open-source blockchain-based satellite network in the world as a case study of applying blockchain theory in designing and implementing satellite systems.

**Keywords:** Blockchain, Space Digital Tokens, Space Industry, and Exploration, SpaceChain

## 1. Introduction

As reported in **[1]** the global space economy in 2018 was worth USD 360 billion. This is a three percent growth in one year. There are two types of space activities: governmental and commercial. Governmental activities include civil applications (such as science exploration and weather forecasting) and military applications (such as imagery and communications). Commercial activities include launching, communications, and satellite imagery **[2].** It is very difficult to imagine our modern life without services such as digital maps, navigation, satellites mobile communication, etc. Such services are mainly based on space technologies. For the modern space systems to sustain the fast technology pace, they need to consider a list of factors mentioned in **[3].** Among these factors comes there two factors of particular interest: using terrestrial computing systems supporting high power and reliable software as well as using applied sciences sustaining information processing and managing complex systems.

In terms of the impact of the space industry on the global economy, there is a huge impact. According to **[4],** the global space economy is estimated between £155 to £190 billion. This would further grow up £400 billion by 2030. The UK space sector has got £13.7 billion turnovers in 2014/15 where it contributes between 6.3% and 7.7% of the global market. The UK uses space technologies in different fields including emergency services, defense, flood response, environmental monitoring, media and other essential functions of the state.

Blockchain is a very promising technology that has proved itself to be successful in many fields such as currency, process management, supply chain, security, managing complex systems, and others. According to [5], blockchain-based technology would contribute by 3.1 trillion USD in the economy by 2030. Since blockchain is designed as a distributed ledger, it will play a major role in managing and controlling data communications between different devices. It also offers capabilities of managing massive patterns of transactions form device-human, device-device, human-human, or human-device.

The basic idea of blockchain is that it is considered a distributed database of transactions with the append-only feature. Given this, blockchain is considered as a strict version of distributed ledgers [6]. In the blockchain-based systems, trust and administrative responsibility are collective between their operators. Blockchain, unlike traditional database architectures, require its participants to perform consensus operations for information to be added to the chains. The security and trust are achieved through cryptographic operations linking a new block to its preceding blocks. This means that blockchain technology is difficult to compromise as it requires a successful attacker to somehow successfully gain access and control many participants and this is theoretically very difficult to achieve.

Utilizing these blockchain properties, they can be very useful techniques/technology to prevent "cyber and physical attacks" against space resources (e.g., spacecraft, Identification management, and data access, etc). Usually, the space industry is collaborative (e.g., multiple parties participate in maintaining and updating of space equipment). Thus, blockchain would a very promising technology enabling such collaborative processes [7]. Also, blockchain would be used to solve a complex problem in secure communication for existing spacecraft. Addressing this problem has been limited because of the relative cost and complexity of satellite ground stations. Operating the spacecraft and its ground station is at risk of availability problems (denial-of-service attacks could be mounted). According to [8], few security attacks have been reported such as the unauthorized access committed at networks controlling spacecraft at NASA JPL. Also, it was reported that U.S. Air Force satellites ware "jammed by commercial equipment easily acquired by the state and nonstate actors".

This paper aims to investigate adopting blockchain theory in the space industry and applications. Also, the paper introduced a new conceptual blockchain framework for proposing new blockchain models that can be used as theoretical solutions to some major challenges in the space industry. Moreover, the paper is ended with discussing SpaceChain as a case study of applying blockchain theory in designing and implementing new satellite systems.

The rest of this paper is organized as follows. Section 2 discusses and explains the theory of blockchain and how the concept of *digital tokens* can be used in several blockchain applications. Section 3 discusses the proposed blockchain framework and presents new conceptual solutions and new blockchain models for some major challenges in the space industry and exploration. Section 4 discusses *SpaceChain*, the first open-source blockchain-based satellite platform in the world as a case study of adopting blockchain theory in designing and implementing the coming new satellite systems. Finally, section 5 presents the conclusion.

## 2. Blockchain Theory and Digital Tokens

Blockchain has emerged as a powerful technology in digital cryptocurrency and online financial transactions since 2008. Bitcoin is the first digital currency depend on Blockchain which invented by an unknown person named Satoshi Nakamoto. The theory of blockchain refers to the possibility of managing an infinite number of transactions within a Peer to Peer network, and stores those transactions in a cryptographically secured data unit called "Blocks" **[9].** As depicted in **Figure 1**, the blockchain-based network consists of many computers (known as miners) working together as peers in a way that the data in the blocks cannot be altered except through a blockchain protocol (or blockchain consensus) **[10].** Moreover, each block in this network has two have values: a unique hash value (or signature) that identifies the identity of this block, and another hash value for its unique parent (i.e. the previous block). The first block in the blockchain is called genesis block which has no parent.

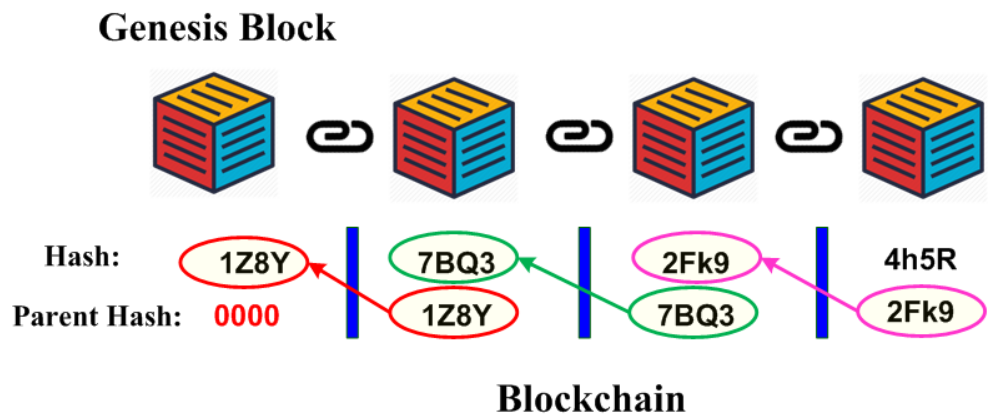

**Figure 1** blockchain as a sequence of hashed blocks

To know how the transactions between the peers are processed and stored into blocks? and, how those blocks are verified then added to the chain? Those questions can be answered in the following subsection.

### 2.1 How Blockchain Works

This subsection gives an overview of the blockchain process. It explains how transactions between the peers are processed and stored on blocks and how those blocks are added to the chain. The blockchain process can be summarized in the following steps.

**Step 1:** a given node creates a specific transaction, for example, sending digital currency to another node through the blockchain network.

**Step 2:** the transaction then broadcasted over the network, which waits to be caught by the miners (the other nodes in the network) who verify the validity of the issued transactions. If this transaction is not grasped by any miner, it hovers as unconfirmed transaction and stored in a queue. It is important to note that it is not a condition to queue all unconfirmed transactions in the same place, but may there are many queues for many unconfirmed transactions. The selected transactions by each miner will be used to form his block as it will be discussed in the next step. So, in this step we have two types of queues, the first type is for the grasped transactions by the miners, and other ones for unconfirmed transactions.

**Step 3:** in this step, each miner starts to form his block (i.e. a nominated informal block) which involves his selected transactions. At this moment, the selected transactions in the formed block still unconfirmed, since the miner should check the validity of the selected transactions. For example, if the transaction is a *digital currency transfer* from user A to user B, the miner should check the history (i.e. previous blocks) of blockchain to confirm that user A 'balance has indeed a sufficient bitcoins to transfer to user B. Moreover, although, every miner can form his block that involves his selected transactions, the same transactions may be selected by different miners in their own formed blocks. On the other hand, if the owner of a specific transaction wants to speed up his transaction process, such as speeding up the bitcoin transfer process, he should offer *a reward* for the miners who add his transaction to their formed blocks. Hence, the miners prioritize the transactions they select according to the better mining reward.

**Step 4:** it is the most important step in the blockchain process since after the miners formed their nominated blocks; they compete with each other in which each miner tries to add his block to the blockchain as a valid one. So, the important question here is, when does a specific block formed by a specific miner become a valid one? Another important consecutive question that rises here also is, how to verify the validity of that block for adding it to the blockchain as a valid and formal block. We can answer these two questions as in the following substeps:

1) Firstly, the blockchain-smart contract has to define and specifies the validity and uniqueness of the newly created blocks using a *hash code*. For example, Proof of Work (PoW) **[11]** specifies that the hash code of a valid block is calculated as in **Equation 1**, and the valid block' hash code must start with a certain number of consecutive zeros (e.g. seven consecutive zeros).

$$BHC = Hash(TD + P + Nonce) \qquad (1)$$

Where 'BHC' is the Block Hash Code, 'TD' is the Transaction Data, 'P' is the hash code of the previous block in the chain, 'Nonce' (**N**umber only **once**) is an arbitrary number that can be used only once for producing the corresponding BHC.

2) Every miner works with his block to produce the right hash code of the valid block that has to be added to the blockchain as a new block. This process is done by solving a complex mathematical algorithm that associated with miner block; this process is called *block mining*. The block mining process requires a lot

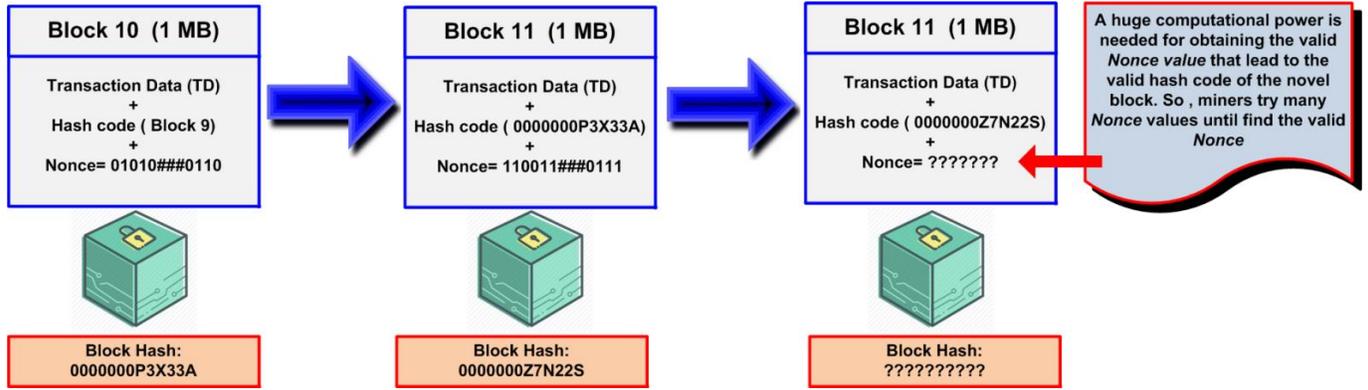

**Figure 2** The blockchain hashing

of computational power for calculating the complex algorithm of the block to get its BHC value.
3) Then, the miners try many times a random number called *Nonce* (Number only once, through the used mathematical algorithm for finding the valid hash code of that block. So every time the *Nonce* value of a block is updated, the block gets a new hash code.
4) The Miners repeat substep 3 of updating the *Nonce* value many times until they randomly get an output string that meets the target block hash code (for instance, a string should start with consecutive seven zeros as in PoW protocol)

**Figure 2** explains how valid blocks are hashed and created in the blockchain.

**Step 5:** when a specific miner succeeded to find the valid hash value for its block first, he then broadcasts this block and its hash code (BHC value) to all the other miners in the network for verifying its validity.

**Step 6:** After broadcasting the new block across the network, all peers (i.e. miners) now verify the validity of the block's hash code, by reversing the hashing function to check if this hash value leads to the corresponding valid *Nonce* as in Equation 2. If it is valid, all peers will confirm and accept its validity and approve that the new block can be added to the blockchain. Then, the blockchain state is updated at all nodes with the added block.

$$Nonce = Hash^{-1}(TD + P + BHC) \qquad (2)$$

## 2.2 *Blockchain as Digital Token Applications*

The blockchain name has often been associated with processing the cryptocurrencies since the appearance of Bitcoin in 2008, but recently, blockchain can be extended more to involve processing cryptographic-tokens (i.e. digital tokens, or digital assets) **[12].** Cryptographic tokens represent programmable valuable assets in any system that can be processed and controlled by a smart contract (or blockchain protocol) and a distributed ledger. For instance, in the healthcare system **[13-16],** drugs can be represented as cryptographic tokens in drug tracking systems, such that drug fraud is one of the most complex challenges in pharmacology. Since blockchain is time-stamped, immutable, and verifiable, drug fraudulent can be detected easily if the drug supply chain process is managed by a Private/Consortium blockchain platform and Proof of drug authenticity protocol.

Also, the digital asset may be programmed as an energy token on the Internet of Energy (IoE) and smart grids **[17][18].** Blockchain can be integrated with Artificial Intelligence mechanisms for specifying the energy consumption patterns in energy sales applications **[19].** Moreover, managing and monitoring digital energy transfer within smart grids is another benefit of blockchain on the internet of energy field **[20][21].** Processing Digital assets using blockchain can be found also in the smart agriculture sector **[22].** It can be used in farm overseeing **[23],** food safety **[24]**, supply chain **[25],** livestock monitoring **[26]** as well as digital crop tracking **[27].**

## 3. Space Challenges and Solutions: Blockchain Framework

Recently, the space industry isn't far from blockchain benefits and its contributions. In 2017, NASA announced a $330.00 grant for developing a blockchain-based spacecraft system **[28].** The grant is provided to Dr. Jin Wei Kocsis for her project Resilient Networking and Computing Paradigm (RNCP) **[29].** RNCP represents the first step by NASA toward blockchain adoption in space applications. Space assets such as satellites, spacecraft, space debris, orbits, and asteroids, can be tokenized as digital tokens and manipulated by blockchain technology **[30].** Converting space assets into digital tokens and handling it using blockchain protocols and smart contracts will add more transactions with space resources from any place in the world. Moreover, blockchain will add more intelligence to spacecraft, where it will be able to 'think' and take crucial decisions without connections with the ground stations.

This motivation leads us to propose blockchain *framework* for providing new conceptual models that explain the strengths of using blockchain in the space industry.

### *3.1. Blockchain Tracking Data Relay Satellite (TDRS) Connections*

In Tracking and Data Relay Satellite (TDRS) system **[31],** blockchain can be used to sync and manage many connection patterns between the TDRS and its follower satellites or between the follower satellites and ground stations network for optimizing user queries from TDRS. For example, users can send a specific image request pattern to TDRS system by entering his request information (e,g, locations, and timeframes) into the blockchain which stored as *user request digital tokens*. Moreover, additional digital tokens such as the transaction session information and uplink data from the ground station to TDRS are also created and recorded in the blockchain as *transaction digital tokens,* which then sent to the ground station network system.

After that, the ground station network forwards the coming data to TDRS in the form of *uplink digital tokens*. Based on specific machine learning techniques, the TDRS can smartly reallocate its follower swarm of satellites based on learning methods to optimize image output and minimize the time and the cost to users' queries. The image output, downlink time to the ground station, process start/completion time, satellite feedbacks to the user are sent to the ground stations network in the form downlink digital tokens, which then recorded in the blockchain system as *TDRS Feedback digital tokens*. Then, user terminals can get the required satellite images after updating the blockchain status with *TDRS Feedback digital tokens* as a new block.

The proposed model in **Figure 3** explains how blockchain can be used to sync and manage these connections when the user sends an image query to the TDRS system by executing the smart contract rules.

## 3.2. Blockchain Satellite Launching Logistics

The satellite launching process is not the responsibility of a unique party. The process is very complex as it involves many participants such as satellite operation center, regulatory authorities, satellite service provider, satellite manufacture, suppliers, launch service provider, and mission control center who can communicate through a P2P network with each other. Moreover, satellite launching logistics require larger funding of research and development; hence, it is not best-suited to be funded by the traditional Venture *Capital* model which is used as a funding standard for small and short-term projects.

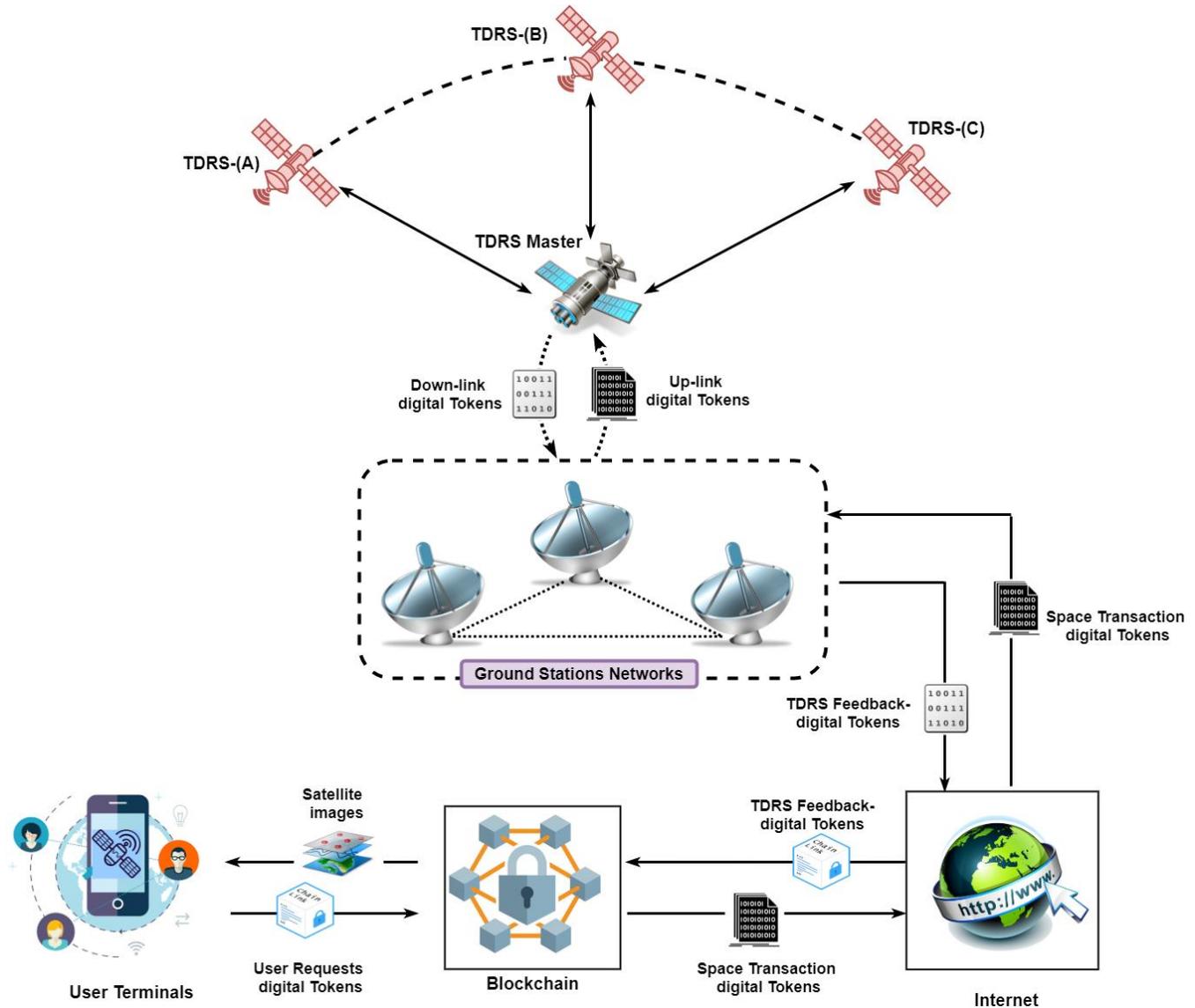

**Figure 3** Blockchain TDRS connection patterns using blockchain as digital tokens

So, new models of crypto-economic funding such as *Initial Coin Offering* [32] have to be adopted for funding long term projects as in satellite mission exploration projects. In the form of funding digital tokens. The major advantage of using crypto-economic funding models in the satellite logistics life cycle is the possibility of utilizing smart contracts as a way of releasing and distributing digital funds in an optimized manner according to the roadmap and planning of launching the satellite logistics process.

So, to mitigate the complexity behind satellite launching logistics and funding consideration, a consortium blockchain is needed, (i.e. a blockchain system that is 'semi-private' and has a controlled miners group, but can work across different organizations ). The complexity of satellite launching logistics process will be mitigated by converting some main assets in satellite launching logistics (e.g. mission requirements, Feasibility results, mission schedules, design models, flight model information, testing results, Real-time and lunching control data ) into digital tokens and processing it by smart contracts.

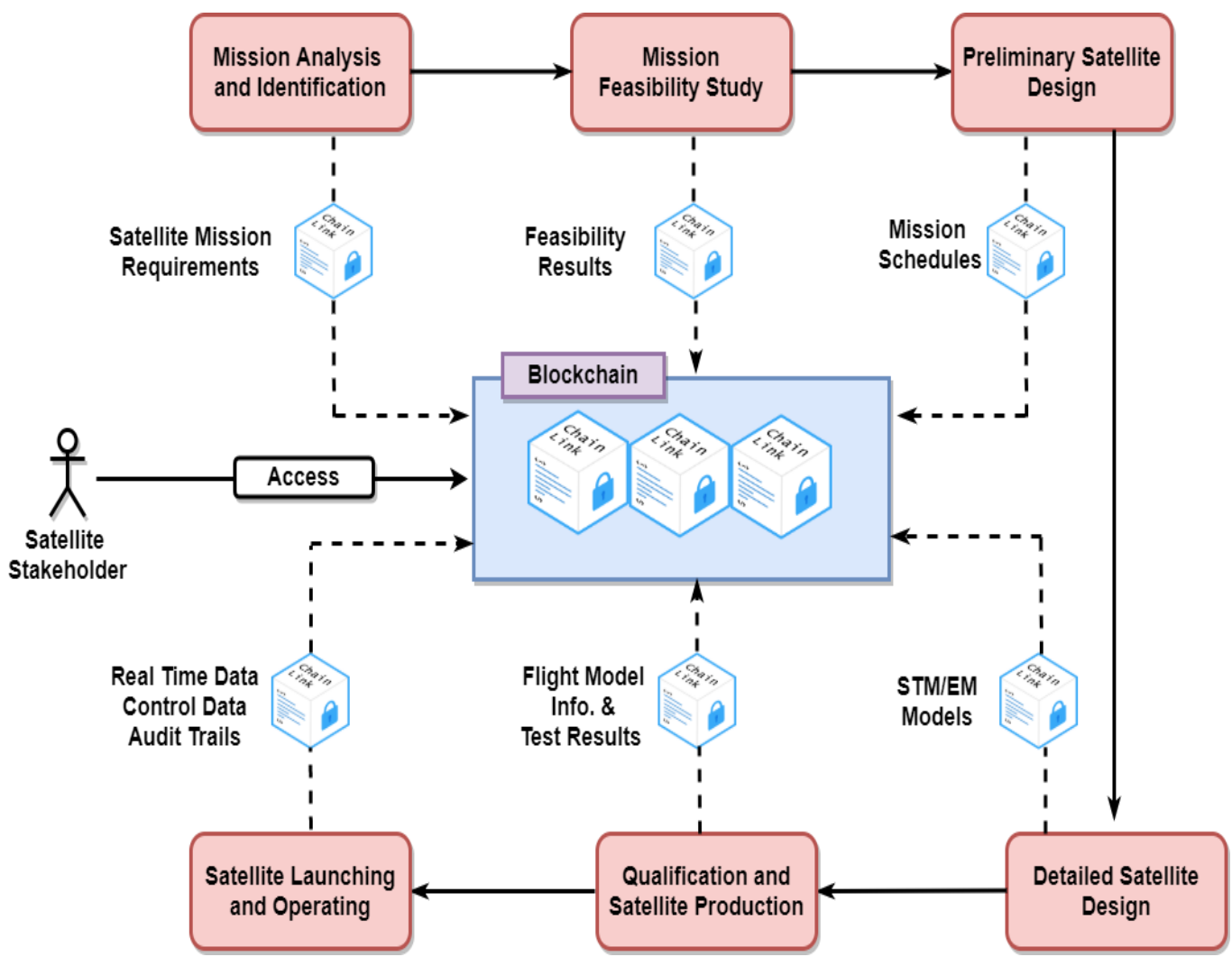

**Figure 4** Blockchain satellite mission data life cycle using a consortium blockchain

**Figure 4** explains a proposed process activity model that explains how the consortium blockchain can be utilized to store, manage, and secure satellite mission data during all phases of launching the satellite life cycle. Each phase can update the blockchain with a new data block pattern that can be accessed by all satellite stakeholders who are authenticated in the consortium blockchain.

In the *mission analysis and identification* phase, the stakeholders conduct many brainstorming sessions to define the requirements of the candidate satellite system. These requirements then verified by the consortium blockchain miners, once they confirm these requirements, the first block are created for storing these requirements and update the blockchain status at all participants in the satellite network.

In the *Mission feasibility study* phase, cost estimations of the satellite mission are outlined, then forwarded to the consortium blockchain to be verified and confirmed by the satellite mission consortium. The confirmed costs and feasibility results then added as a new block in the blockchain and trigger a new blockchain update in the network.

In the *Preliminary Satellite Design* Phase, new satellite components interfaces, and mission schedules are defined and verified by the blockchain administrators. The confirmed design schedule data are stored in a novel block in the blockchain and trigger a new update in the blockchain network.

In the *Detailed Design* phase, two versions of the satellite are designed: the Structural and Thermal Model (STM) and the Engineering Model (EM). All details and metadata of those models are verified and stored as a new block in the blockchain system.

In the *Qualification and Satellite Production* Phase, metadata about the produced Flight Model (FM) is produced; Environmental tests, thermal vacuum tests, Vibration tests, and acoustic tests are applied. All FM metadata and test results are verified and stored as a novel block in the blockchain.

Finally, in *Satellite Launching and Operating* phase, launching control data, satellite tracking data, real-time data, and audit trails are also verified by the satellite mission consortium and stored as a new block in the blockchain.

## 3.3. Orbital Assets Tokenization using Blockchain

We can define orbital asset tokenization as the process of transforming the major space entities (e.g. satellites, orbit vectors, space debris, asteroids, spacecraft, astronauts, etc.) into digital tokens that can be recorded, and processed by a blockchain system. Tokenizing space assets would enable space stakeholders to monitor and track the intended orbital tokens using a smart contract written into the blockchain network in a transparent and evident way. **Figure 5** depicts a proposed model that explains the process of tokenizing orbital assets and processing these tokens in the blockchain.

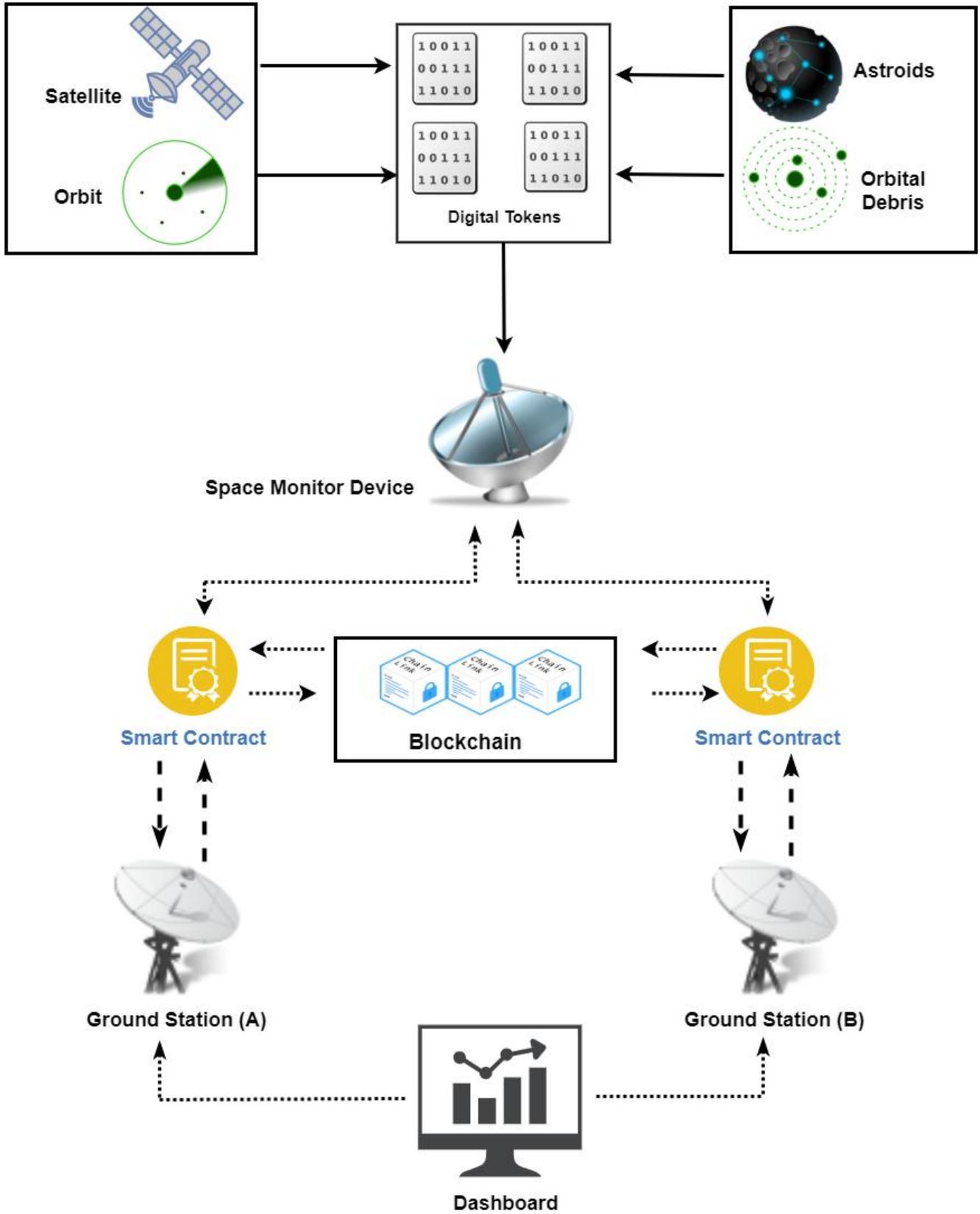

**Figure 5** *Orbital assets tokenization using blockchain process*

Orbital assets, such as orbits itself, satellites, asteroids, and orbital debris can be represented as digital tokens that specify important orbital data to space stakeholders. The space monitor device can work also as a transceiver to the transformed digital tokens. After that, a variety of smart contracts installed on different ground stations can process the delivered orbital tokens and produce *space decision tokens* which capsulated as a new added block to the public blockchain. All space stakeholders become able to access the newly added blocks through the connected dashboard.

The main advantages of tokenizing space assets by blockchain technology can be described as follows:

1) *Greater Control over Space Assets:* tokenizing space assets into secure digital tokens will enhance tracking, processing, verifying, and securing all space behaviors triggered by space objects. Moreover, will manipulate them through a consortium blockchain system based on smart contracts. Also, the intended stakeholders of tokenized space asset data are in greater control, in terms of granting and revoking access to their satellites and spacecraft, as well as data sharing from and to space.
2) *Faster transactions and Automated Compliance:* Compliance and fast response is a major factor in space asset management. Smart contracts based on the blockchain could prove beneficial in automating variety of space transaction patterns such as satellite-to-satellite (S2S), Ground station-to- satellite (G2S) and vice versa, user- to-satellite (U2S) and vice versa, Ground station to another one (G2G) and Astronaut-to Astronaut (A2A), etc. The long time that a space message takes to reach the ground stations will be eliminated by using blockchain systems, where, blockchain can manage up/downlink transactions between satellites and ground stations in real-time.
3) *Security:* Blockchain-enabled decentralization for space networks eliminates the risks involved in the traditional single-point-access model of space communications. Moreover, space assets on Blockchain are cryptographically encrypted and making them tamper-proof. Also, blockchain enables the space companies to track and monitor their satellites and spacecraft in more transparency and control. Blockchain will enable space agencies to detect any attack on a satellite system that seeks to tamper the satellite software system using space bots **[33].**

### 3.4. Securing Satellite Swarms communications

Before we explain how blockchain can be utilized for securing satellite swarms communications, we should explain the communication patterns between satellites and blockchain system. These patterns can be classified as four communication models as stated in **[34]:**

1) A Satellite/spacecraft works as a blockchain node within the blockchain network
2) A Satellite/spacecraft works as a validator (i.e. a miner) node within the blockchain network
3) A satellite/spacecraft read from the blockchain
4) A satellite request a specific transactional data to be stored in the blockchain

**Figure 6 (a,b,c,d)** shows a proposed model that describes four communication patterns between satellite and blockchain-satellite network.

Developing swarms of multi-sensor satellite systems for earth and space explorations require establishing decentralized platforms for managing and securing space communication patterns efficiently.

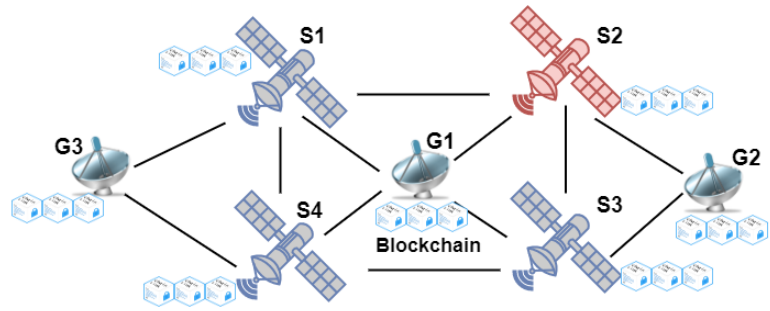
(a) the satellite S1 works as a blockchain node

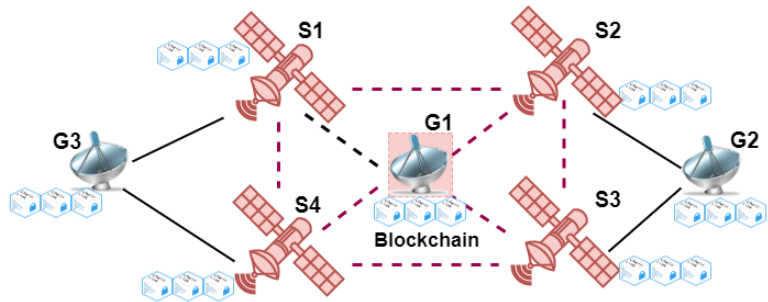
(b) the satellites S1, S2, S3, S4, and G1 works as set of miners in the blockchain network system

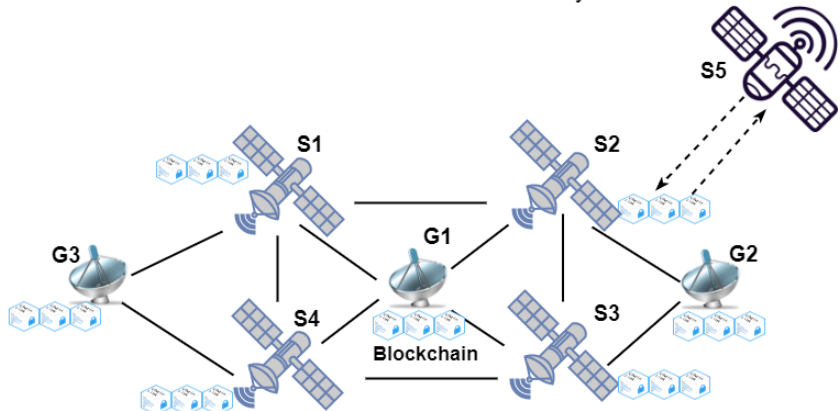
(c) the satellite S5 Reads data from the blockchain through the connection with the satellite S2

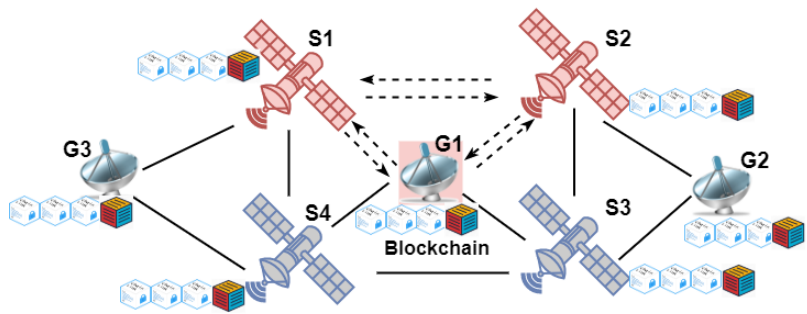
(d) the satellite S1, S2 establish new connection through the ground station G1 and the blockchain is updated with a new block

**Figure 6** Four communication patterns between satellites and blockchain-satellite network

The communications between ground stations and satellite swarms require decentralized tracking and monitoring of active and inactive resident space objects. Also, it requires assessing the space environment through a network of multi and heterogonous sensors of satellite swarms in different orbits. Blockchain can also be utilized for securing satellites swarms communications and authenticating space transactions between those swarms and ground stations **[35].**

Blockchain can be utilized for creating virtual trusted space zones in which the satellite swarms could identify and update each other in a trustless network environment. For instance, when a specific satellite is threatened with a space debris collision within its orbit, it will update all satellites in the same swarm (i.e.in the same orbit) with the new information. This update is distributed as a digital token within the satellite swarm based on a blockchain network system. On the other side, a satellite swarm in LEO orbit may establish another connection with another MEO/GEO satellite swarm for providing a specific service to the ground stations network. This connection pattern also is managed by a blockchain network.

**Figure 7** depicts a proposed model that shows how blockchain can be utilized for managing and securing the connection patterns between three satellite swarms that represented as three virtual zones space. A specific satellite in Zone A can establish a connection with another satellite in zone B through the onboard multi-sensor system in each one. Then, the feedback of this communication is represented as a new space digital token that encapsulated in a newly added block to the blockchain system. The major security operations that blockchain can provide for securing the communications of satellite swarms (or virtual zones) can be discussed in the following sub-subsections.

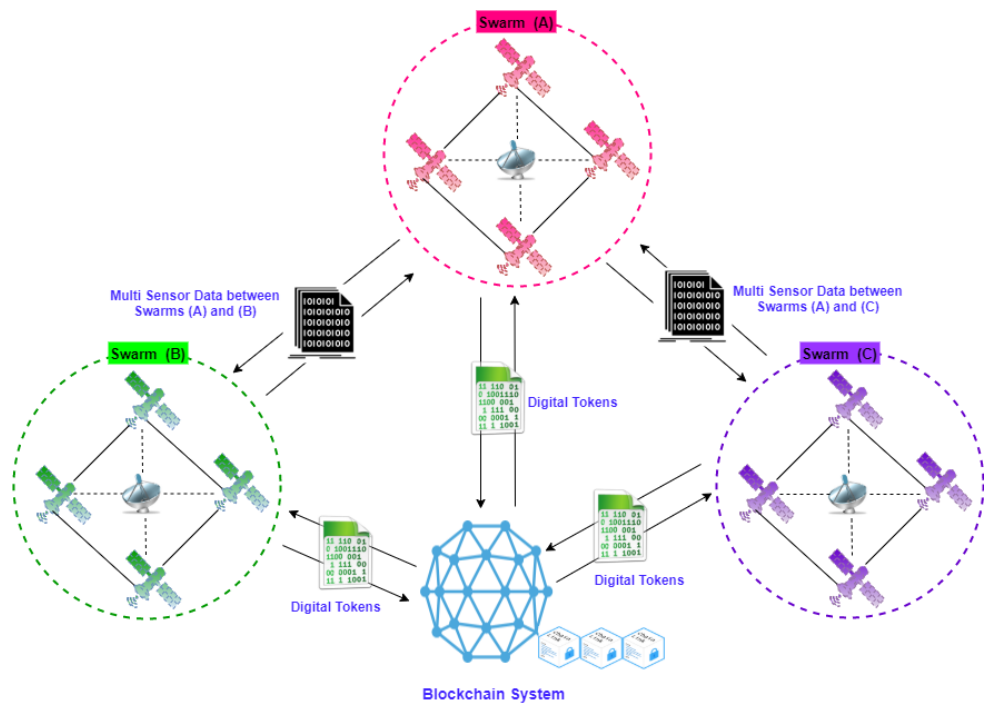

**Figure 7** managing and securing satellite swarm connections using blockchain

1) <u>Multi-Factor Authentication (MFA) in Space</u>

In space communication theme, Multi-Factor Authentication (MFA) can be redefined as a security technique that verifies satellite's identity, ground station's identity, or communication pattern validity by requiring multiple security proofs. For example, when a satellite A request a specific connection with a satellite B within a satellite swarm, the satellite B ask the satellite A to send the last block's *Nonce code* in the blockchain. The Nonce code is used to prove the satellite's membership in the swarm, and then the Nonce code is verified by the satellite B for establishing the communication. After ending the connection between two satellites, a new block with a new Nonce code is issued, and then the new block is verified by the miners (e.g. satellites and ground stations in the blockchain network) for proofing the validity of connection and adds the new block to the blockchain system. **Figure 8** depicts a proposed model that explains how blockchain can be used for authenticating the connection between two satellites using the MFA technique.

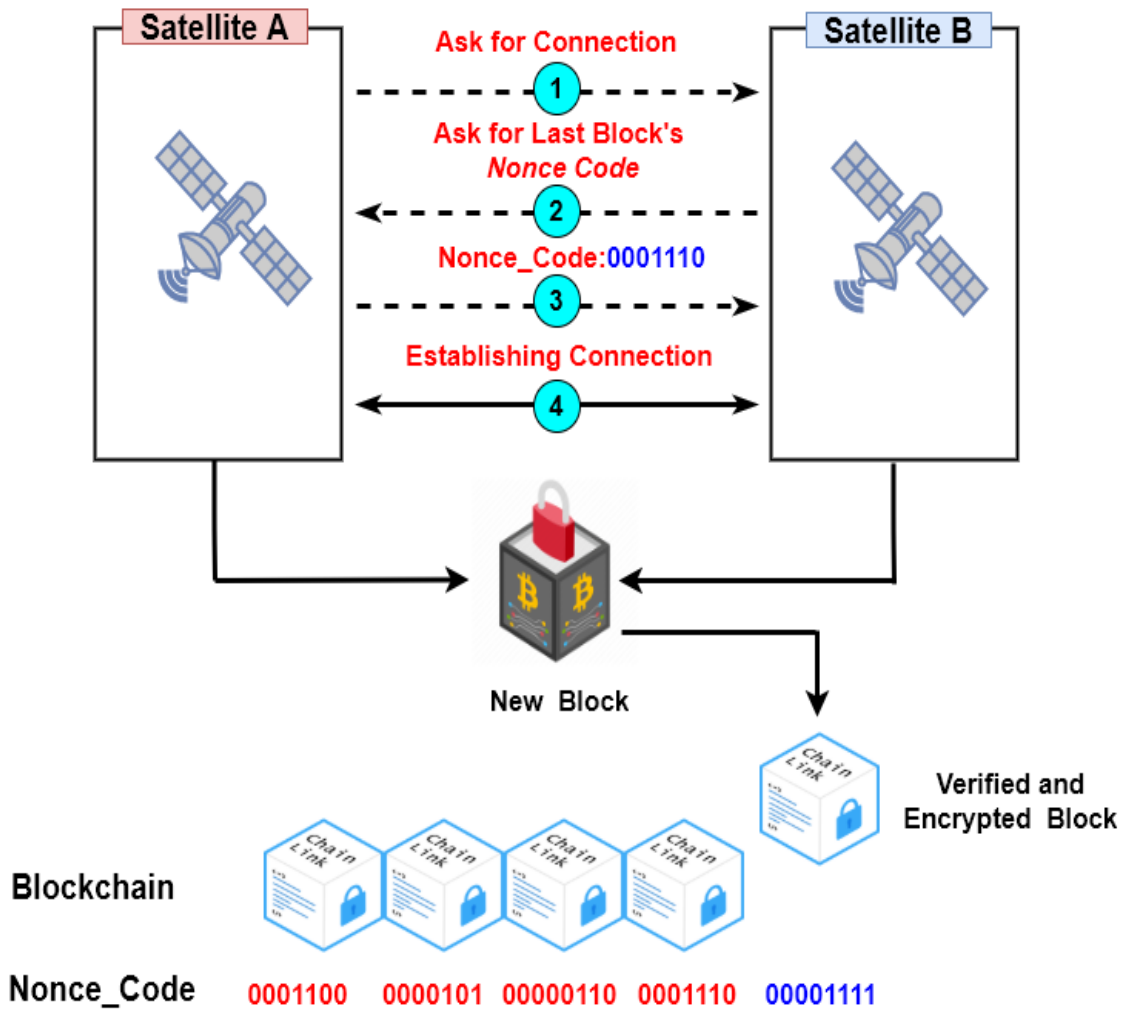

**Figure 8** Authenticating satellite communications using blockchain-based MFA technique

2) Creating secure Virtual Zones in Space

In this scenario, blockchain can also play another important role in securing space communication by creating secure virtual zones in space. Space Virtual zone refers to a swarm of satellites in LEO, MEO, or GEO orbits that can be managed by one or more ground stations called *Zone master*. Creating a specific virtual trusted zone for a swarm of satellites is mapped to the first block in the blockchain which called the *Genesis Block*. So the rising question here, how virtual zones are created and how blockchain can be utilized to secure them?

Let us go with the first question, creating virtual zones depends on the *zone master* which manages and controls a swarm of satellites within the same orbit. The zone master specifies the identity of each satellite in the swarm by issuing a *Virtual ID* for each one. Then, the zone master records these IDs of satellites as well as the *Virtual Zone ID* in the *genesis block* of the blockchain. Moreover, defining all requirements, conditions, rules of the future satellites' join to the created zone are specified in a smart contract (i.e. blockchain protocol) which executed by the *virtual zone master*.

**Figure 9** depicts a proposed model that explains three virtual zones for three swarms of satellites in LEO, GEO, and MEO orbits. The three virtual zones are created and managed by the corresponding virtual zone master. The zone master issues the Zone ID, and authenticates its satellites using virtual IDs, and then the virtual zone id and satellites virtual ids within each zone are registered in the first block in the blockchain (i.e. genesis block).

The second question is about how blockchain can secure the created virtual zones and protect its swarm of satellites against space attacks from outsider satellites. As we mentioned above that all requirements and registration rules of adding new satellite to a specific virtual zone are specified in a blockchain smart contract which executed by the zone master. So for example, if a new satellite asks to join a specific virtual zone, its *join request* must be verified by all satellites in the virtual zone for proofing the validity of this new space communication. The verification algorithm is executed according to the specified join rules defined in the smart contract. If the transaction is valid and the identity of the new satellite is verified, the zone master issues new virtual id to this satellite and add it to the zone and register this new transaction as a new block in the blockchain. Otherwise, the domain master rejects this communication and detect this satellite as an intruder one as depicted in the proposed method in **Figure 10**

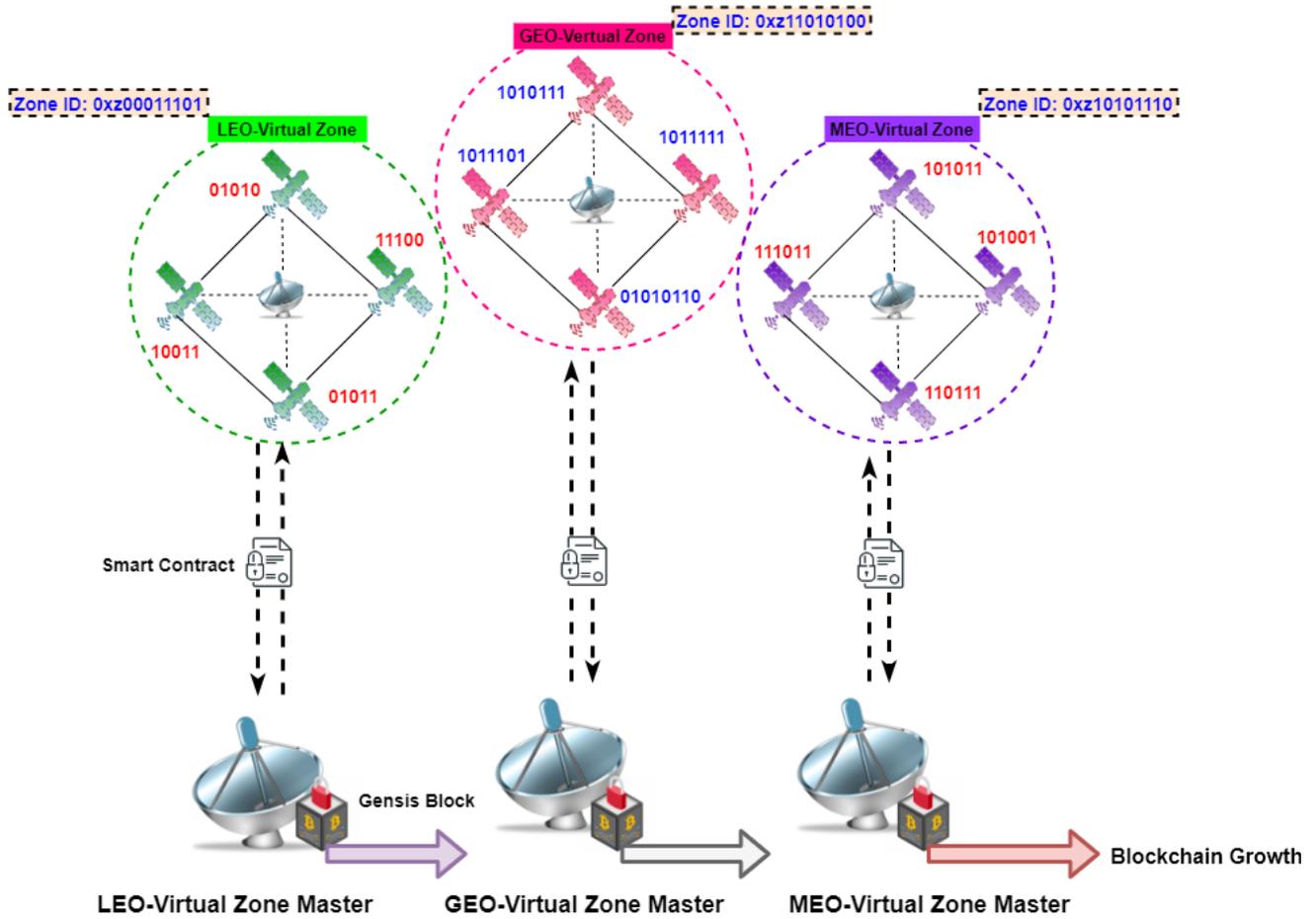

**Figure 9** Creating virtual zones using blockchain

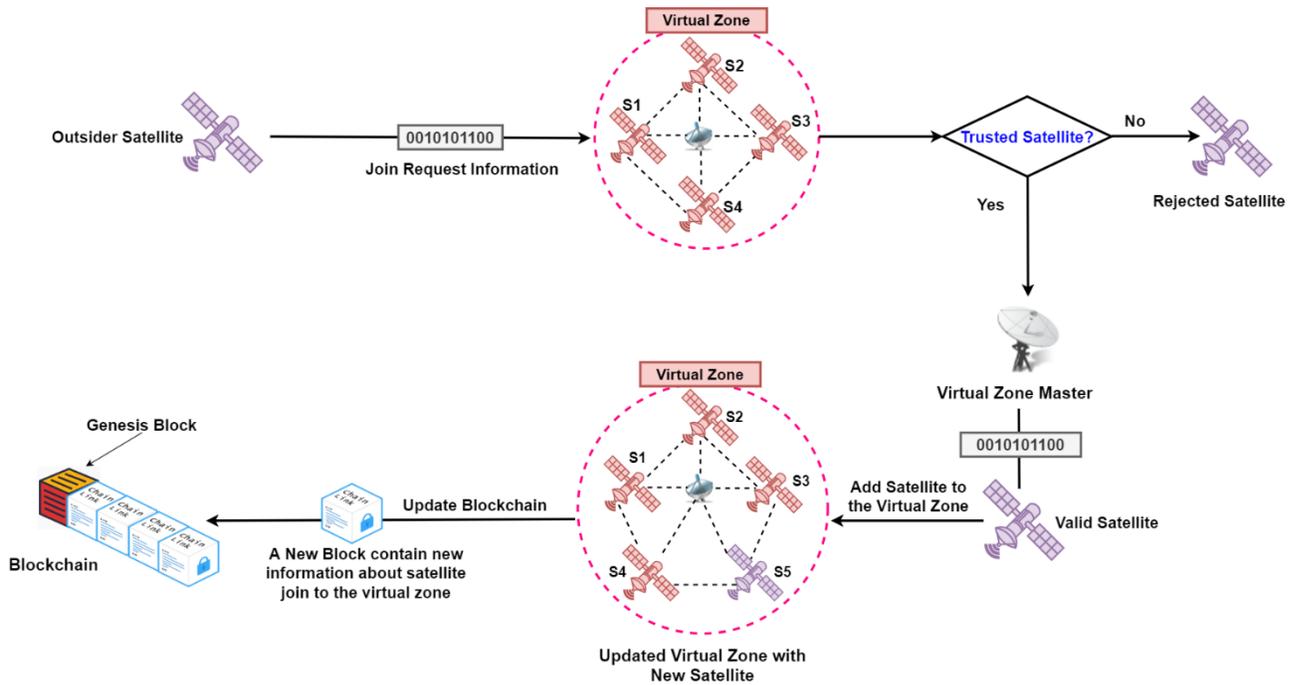

**Figure 10** verifying satellite connections to virtual zones using blockchain

3. Protecting Virtual Zones against Space Debris

Space debris is a very big challenge and represents a serious threat against satellite motion within its orbit **[36]**. The collision between a satellite and space debris can crash the satellite or at least deviates it far away from its right orbit. The decentralized design of blockchain can be also utilized for tracking space debris motions and sharing its orbital data such as sizes, speed, altitudes, motion directions, and other attributes among the trusted satellites in a specific satellites swarm. Space debris-orbital data can be shared between satellites swarm in the form of orbital *digital tokens*, and stored on the blockchain. For example, **Figure 11** depicts a proposed model which explains how blockchain is updated when a specific satellite within the swarm sense close debris move toward space virtual zone. When the satellite sense close debris, it sends the sensed data to the *zone master* which uses it for estimating the corresponding smart maneuvers algorithms and other collision avoidance decisions for each satellite in the swarm **[37].** After that, the maneuvers data is capsulated as a new block which added to the blockchain system. According to the new data in the shared block, each satellite will update its orbital status to avoid the collision with the detected space debris.

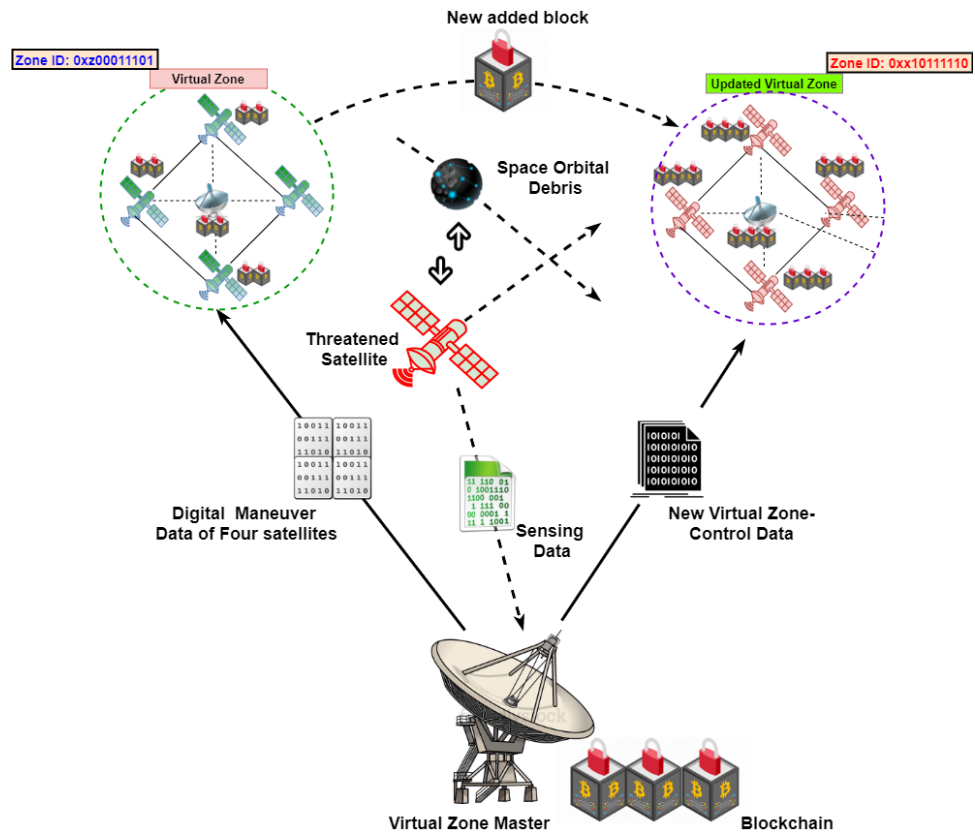

**Figure 11** Protecting virtual zones against orbital space debris using blockchain

4. **SpaceChain: A Case Study**

Founded in 2017 by the SpaceChain team which led by Zee Zheng as CEO (The chief Executive Officer), and the former Bitcoin core developer and the blockchain pioneer Jeff Garzik as CTO (The chief Technology Officer) **[38].**

SpaceChain is defined as a decentralized global space platform that integrates space and blockchain technologies to establish the first open-source blockchain-based satellite network in the world. SpaceChain will permit users to develop and run decentralized applications in space based on an open-source and real-time operating system which will be available to anyone, anywhere in the world. SpaceChain's application is similar to a mobile Apps development platform. With blockchain software, SpaceChain provides an application sandbox or firewall so that immutable and independent space Apps may be uploaded to the satellite as the same as apps on a mobile application Store **[39].**

Streams of space data patterns will be secure and protected through proven blockchain cryptography. In doing so, the vision of SpaceChain is to allow global communities anywhere in the world to communicate access and collaborate in space without any barriers. In February 2018, SpaceChain released and launched its first SpaceChain OS blockchain node into space **[38]**

*4.1. SpaceChain Architectural Design Model*

The SpaceChain architecture is designed as a four-layered model as in **[38].** The SpaceChain design layers can be explained as follows:

1) **Hardware Layer:** the hardware layer involves the *blockchain payload*. It is responsible for processing the space applications securely on the satellite through the SpaceChain OS. In the future, blockchain payload will continue to be updated to meet the future of blockchain-based space applications. Recently, there are more than 320 leading companies and 29 universities participating in the RISC-V open-source project **[40].** SpaceChain is upgrading its blockchain payload processor under the open-source RISC-V ISA (Instruction-Set Architecture) processor standard to run on its OS. The blockchain payload will be updated in the performance and reliability of SpaceChain through the following features:
   - *Microcontroller:* The hardware will be upgraded by combining CPU (Control Processing Unit), DSP (Digital Signal Processing), and other mixed-signal functionality on a single device.
   - *Operating System:* The On-Board Computer (OBC) will work on a default SpaceChain operating system
   - *$I^2C$ Interfaces.* The OBC will have $I^2C$ buses that allow bidirectional data transfer between masters and slaves without data corruption. the devices with different bit rates can be communicated via one serial bus which used as a handshake mechanism to suspend and resume serial transfer based on Serial clock Synchronization technique
   - *CAN Interface:* Controller Area Network (CAN) bus is one of the main OBC interfaces that allow communication between hardware subsystems through a serial communication protocol that supports distributed real-time control with a high level of security.

2) **Operating System Layer:** this is the most important layer in the SpaceChain architectural model. Because of the variety in space communication patterns and satellite systems, space communications needs a stable and versatile operating system. In response to this need, SpaceChain Organization developed a SpaceChain Operating System which is the first blockchain operating system designed for satellite and spacecraft communications. With SpaceChain OS, developers become able to design space applications and install them into satellites. Strength with the SpaceChain OS is sandboxing which immunizes and protects all space applications against confusion and interferences.

The main components of SpaceChain OS can be categorized into three components:

- *Satellite Hardware Drivers:* involves the firmware that manages and controls satellite devices such as power systems, sensing systems, transceivers, antennas, etc.
- *Real-time Operating System (RTOS).* RTOS is intended to serve real-time space applications that process space data streams as it comes in from other satellites, ground stations, or users, typically without any buffer delays. SylixOS is the most open source Real-Time Operating System which used in SpaceChain OS. In the satellite platform. SylixOS can be used to manage satellite hardware drivers such as batteries, sensors, transceivers, memory, antennas, etc . and in the Blockchain-satellite payload, it can be used to provide standard API interfaces and various system features for space applications.
- *SpaceChain Software Development Kits (SDK).* The SDK facilitates the possibility of developing modular space applications and provides a modular development environment that is similar to JavaScript. Moreover, the open-source feature allows and encourages people to join the SpaceChain network easily. POSIX (Portable Operating System Interface) is a common example of SpaceChain SDK.

3) **Software and system Library Layer.** This layer involves some software libraries that can be used for developing and processing blockchain-based space applications. For example, the Libsnappy library can be used for high speed and reasonable compression/decompression space data. Moreover, Libcrypto++ is another open-source library that can be used for cryptographic algorithms. Libleveldb is an additional open-source library that can be used for fast key-value storage that provides an ordered mapping from string keys to string values, so it can be used for storing blockchain metadata.

4) **Blockchain Applications layer.** This is the top layer of SpaceChain system that can be divided into two sublayers in bottom-up: Qtum blockchain sublayer, and Decentralized Apps sublayer:

- • *Qtum blockchain:* Qtum is an open-sourced public blockchain platform. It depends on UTXO (Unspent Transaction Output ) as a secure digital currency which used to transfer UTXO coin to the receiver's Public Key. Qtum is based on Proof of Stake (PoS) and Decentralized Governance Protocol (DGP) which allowing specific blockchain settings to be customized and modified by making use of smart contracts **[41].**
  Qtum-blockchain can provide some important services to the top-decentralized layer, such as:
    - Providing a sandbox to ensure the security and interference-free of space applications processing
    - Providing a compatible and flexible environment for running multiple virtual machines such as Ethereum Virtual Machine (EVM ) and the revolutionary Virtual Machine ( x86 VM.) to process Qtum smart contracts.
    - Providing compatibility of connecting Qtum with other public blockchain platforms and executing its smart contracts in a valid form
- • *Decentralized Blockchain Applications*. This sublayer presents some distributed applications that can be installed in the satellite, such as, *cloud computing applications* for space data recovery. *Cryptography Applications* for encrypting/decrypting space data sent from/to the satellite. *Multi-Wallet Ecosystem Applications* which is a crypto wallet applications that can support different users' needs across different operating systems.

Moreover, this sublayer can involve other local applications such as IoT Apps, Education Apps, and chat apps.

*4.2 SpaceChain Satellite Systems*

The SpaceChain satellite system consists of five main subsystems; satellites, user terminal, ground station, cloud services, and core network

1) **Blockchain Satellite System.** Blockchain-based Satellites consists of three parts: platform; payloads, and Integrated Management Unit (IMU). The platform is responsible for supporting the payload in space, the payload works to realize all the specified satellite functions, and Integrated Management Unit (IMU) is the core of the satellite control components, responsible for coordination and control of all equipment to realize TT&C function, attitude/orbit control, auxiliary management of payload and information management. The main components of the platform and payloads can be summarized as in Table 1. And Fig 12 depicts the main design model of the blockchain satellite system.

The main functions of blockchain-based satellite system can be categorized as follows:

- Two-way communication with the user terminal
- Transfer data to an earth station, update TT&C data from an earth station
- Communication between satellites makes the entire system in real-time
- Processing normal applications/blockchain
- Handling smart contracts in the earth station
- Receive data from a constellation and send it to the cloud service
- Send OS update code/application code to satellites
- TT & C function

-
2) **User Terminals:** The User terminal is the interface between the SpaceChain system and the end-user. It can execute the user's commands and receive data that user's demand and process applications and smart contracts. The user can be human, vehicle, machine or plant, etc. The main functions of the user terminal can be categorized as follows

- The interface between the end-user and SpaceChain system
- Achieve user's command/receive data which user demand
- Processing normal/Blockchain applications including smart contracts
- Users can be anyone or anything who needs and wants to use this network such as human/animal/vehicle/machine/plant etc.

| Table 1: Space chain satellite system components | | |
|---|---|---|
| **Satellite Platform** | | |
| **Component** | **Function** | |
| Main Structure & Mechanism | Mechanic support for all the equipment | |
| Integrated Management Unit | Processing attitude/orbit/thermal control program to maintain the attitude/orbit state of satellite | |
| Sensors | Collect key data to support attitude/orbit/thermal control program; provide necessary data to the payload | |
| Actuators | Execute the order from attitude/orbit/thermal control program | |
| Solar Panel & Battery System | Provide electricity power to all the equipment and payload | |
| TT&C System | Send monitoring data and remote sensing data to the ground station, and receive and execute the order from the ground station | |
| **Payload** | | |
| Payload | Component | Function |
| Blockchain Payload | High-Performance Chip (PSOC) | Provide enough to calculate the ability to support OS and apps |
| | Memories | Data storage/software code storage / Blockchain account book storage |
| | Open source operation system | Provide an OS to support smart contracts and all applications |
| Communication Payload | Satellite user terminal communication payload an antenna | Bi-direction communication with user terminal |
| | Inter-satellite communication payload and antenna | Real time communication |
| | Satellite-ground station communication payload and antenna | Data transmission to the ground station, OS update/applications update |
| **Control Unit** | | |
| Integrated Management Unit (IMU) | TT&C | It receives data directly from TT&C package and executive orders and sends telemetry data directly to the transponder to send to the ground of the earth station |
| | Attitude/Orbit Control | Collect data from all position sensors, implement a position/orbit monitoring program, and control a position/orbit motor |
| | Auxiliary Management of Payload | Receives data and order from the Blockchain earth station to control the payload of communications |

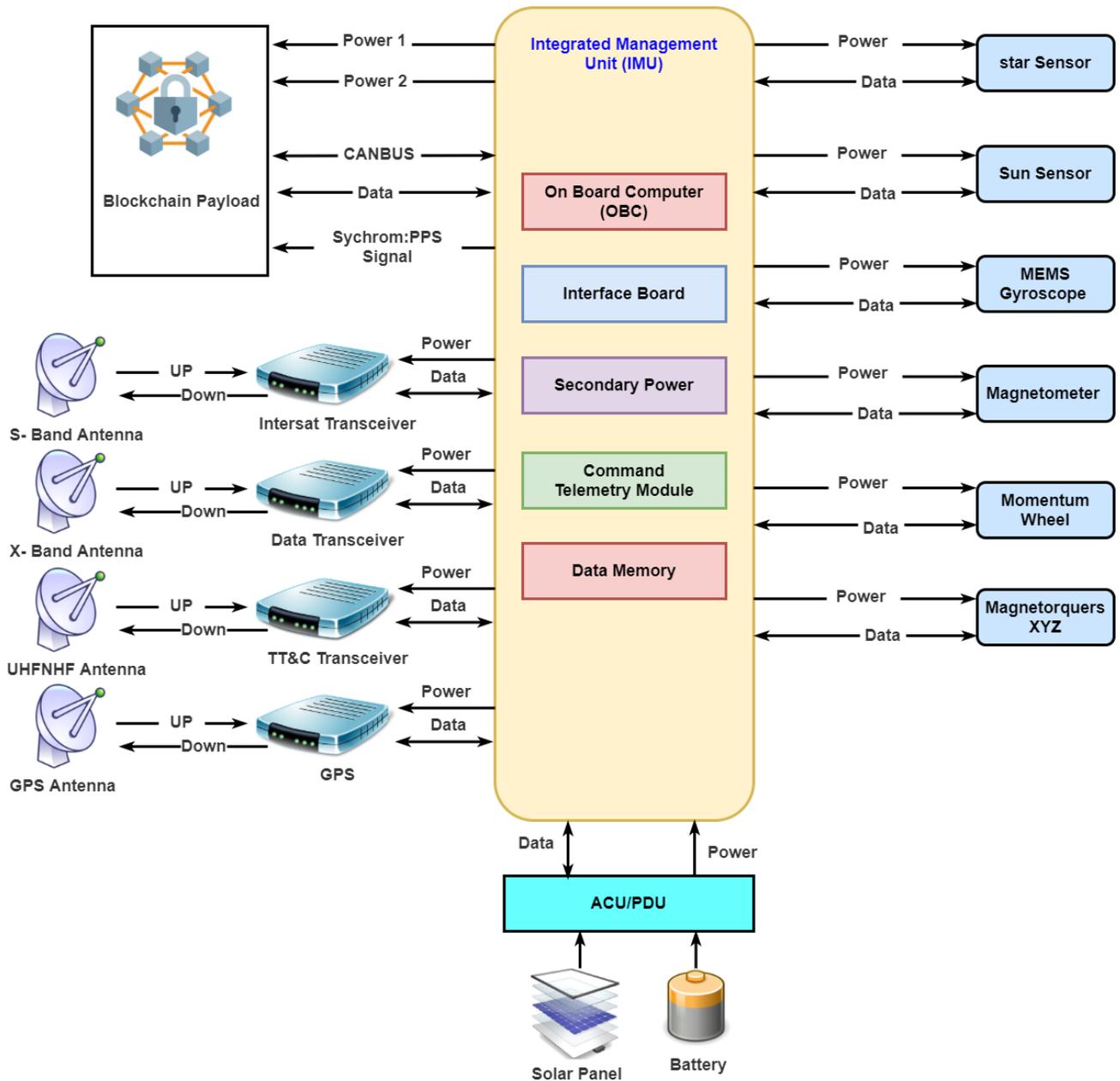

**Figure 12** the architectural design model of Blockchain satellite system

3) **Cloud service** is responsible for assembling data processing and decentralizing cooperative communications based on blockchain technology.
4) **Core Network** is responsible for connecting all ground station systems and linking the SpaceChain network to other networks (caber\cellular networks etc.). So, the core network makes the SpaceChain system a whole network and to be an interface between another network system, like cellular network/cable network, etc.
5) **Ground Station** is responsible for the feeder link between satellites, user terminal management, channel management and operation control; as well as the operational management of the maintenance system. The ground station includes the TT&C system and operation control system.:

   - *Operation control system.* It is responsible for data acquisition gateway station, data acquisition center, user terminal Administration, and user service system. The system is responsible for preparing the satellite mission plan and completing the transmission management for the user station responsible for satellite monitoring. This system is also responsible for receiving, processing and distributing user station data.
- **TT&C System:** The TT&C system combines VHF/UHF frequency measurement and GPS/BD - 2 auxiliary communication protocols, isolating up/downlink by TDD, using rod form as TT&C antenna, realizing auxiliary orbit measurement by GPS/BD antenna and receiver.

As a summary, SpaceChain aims to get the blockchain technology more involved by establishing satellite system models based on digital tokens. Space digital tokens will act like tickets that need to be bought to upload and execute space-based applications through blockchain satellites and store data onto the SpaceChain network. This new space technology encourages more space stakeholders to embrace and contribute to the SpaceChain ecosystem.

## 5. Conclusions

The current study aimed to examine the effectiveness of utilizing blockchain in the space industry and satellite communication. This study introduced a conceptual framework for investigating the main contributions of blockchain technology in space missions. The qualitative findings of this investigation showed that tokenizing space assets in the form of digital tokens and processing it using blockchain platforms can be used to solve big challenges in the space industry. Moreover, new blockchain conceptual process models have been proposed for mitigating some major space industry challenges. The new blockchain models explaining how blockchain can be used for managing satellite mission data during Satellite Life Cycle as well as how blockchain features can be utilized for securing satellites swarms communications through creating secure virtual zones in the space. Moreover, the research ended with studying and analyzing the SpaceChain, the first open-source blockchain-based satellite network in the world as a case study of utilizing blockchain in building satellite systems.

This research appears to be one of the first attempts to thoroughly investigate tokenizing space assets and processing them as digital tokens through the blockchain platform. Although this research is one of the first attempts towards adopting blockchain technology for solving various challenges in the space industry, several questions remain to be answered, and more broadly, research efforts are also needed for conducting experimental studies and simulating the functionality of the proposed blockchain models in this study.

**Acknowledgment:** This research was supported by Scientific Research Group in Egypt (SRGE) and This work is supported by the Academy of Scientific Research & Technology (ASRT), Egypt and coordinated by National Authority for Remote Sensing and Space Sciences (NARSS) under the TEDDSAT Project Grant.

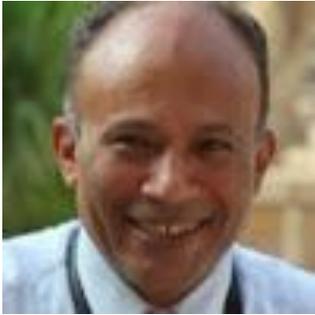

**Dr Aboul Ella Hassanien** is the Founder and Head of the Egyptian Scientific Research Group (SRGE) and a Professor of Information Technology at the Faculty of Computer and Artificial Intelligence, Cairo University. Professor Hassanien has more than 1000 scientific research papers published in prestigious international journals and over 50 books covering such diverse topics as data mining, medical images, intelligent systems, social networks and smart environment. His other research areas include computational intelligence, medical image analysis, space sciences and telemetry mining. Prof. Hassanien won several awards including the Best Researcher of the Youth Award of Astronomy and Geophysics of the National Research Institute, Academy of Scientific Research (Egypt, 1990). He was also granted a scientific excellence award in humanities from the University of Kuwait for the 2004 Award, and received the superiority of scientific in technology - University Award (Cairo University, 2013). Also He honored in Egypt as the best researcher in Cairo University in 2013. He was also received the Islamic Educational, Scientific and Cultural Organization (ISESCO) prize on Technology (2014) and received the state Award for excellence in engineering sciences 2015. Dr. Hassanien holds the Medal of Sciences and Arts from the first class from President of Egypt "Abdel Fatah Al-Sissy". In 2019, Professor Hassanien received Scopus award for his meritorious research contribution in the field of Computer Science.

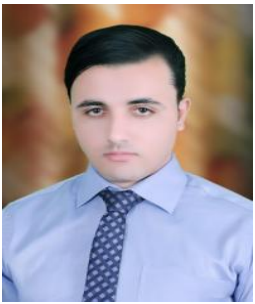

Dr. Mohamed Torky is a senior member of the Scientific Research Group in Egypt (SRGE). works as a Visiting Assistant Professor of Information Technology at faculty of Computer and Information Systems-Islamic University in Medinah - KSA. He worked as Assistant Professor of Computer Science at Higher Institute of Computer Science and Information Systems - Science and Culture City Academy - 6 October City- Giza. His research interests are: Computer and Information Security, Blockchain Technology, Space Science and Satellites, Petri Net-based applications, Graph Theory-based applications, and Automata Theory-based applications.

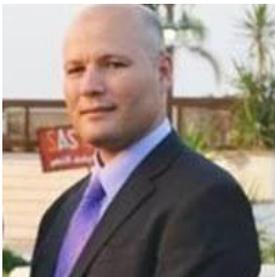

Tarek Gaber is a Lecturer at University of Salford, UK. Dr. Gaber received his PhD in Computer Science from the University of Manchester in 2012. He has worked in many universities including Faculty of Computers and Informatics, Suez Canal University, Faculty of Computers and Information Sciences, Ain Shams University, and the School of Computer Science, University of Manchester, Manchester, UK. He had a postdoctoral position at the Faculty of Electrical Engineering and Computer Science, VSB Technical University of Ostrava, Ostrava, Czech Republic. He is also a senior member of The Scientific Research Group in Egypt (SRGE). He has served as a co-chair and PC member in many international conferences and reviewed many scientific papers and participated in many scientific events (national/international conferences and workshops). He has more than 50 publications in international journals, conferences, and book chapters. In addition, he has 4 edited books.